\documentclass[aps,prl,twocolumn,superscriptaddress,floatfix]{revtex4-2}
\usepackage{amsmath,amssymb}
\usepackage{graphicx}
\usepackage[colorlinks=true,citecolor=blue,linkcolor=blue,urlcolor=blue]{hyperref}
\usepackage{physics}
\usepackage{braket}
\usepackage{xcolor}
\usepackage{multirow}
\usepackage{float}
\usepackage{booktabs}
\newcommand{\GeV}{$\rm{GeV}$}

\begin{document}

\title{First determination of \texorpdfstring{\(V_{cs,cd}\)}{Vcs, Vcd} from inclusive $D$ meson decays}

\author{Kang-Kang Shao}
\email{shaokk25@hust.edu.cn}
\affiliation{School of physics, Huazhong University of Science and Technology, Wuhan 430074, China}
\affiliation{Frontiers Science Center for Rare Isotopes, and School of Nuclear Science and Technology, Lanzhou University, Lanzhou 730000, China}

\author{Hai-Long Feng}
\affiliation{School of physics, Huazhong University of Science and Technology, Wuhan 430074, China}

\author{Xue-Yin Liu}
\affiliation{Wuhan University, Wuhan 430072, China}

\author{Qin Qin}
\email{qqin@hust.edu.cn}
\affiliation{School of physics, Huazhong University of Science and Technology, Wuhan 430074, China}

\author{Liang Sun}
\affiliation{Wuhan University, Wuhan 430072, China}

\author{Fu-Sheng Yu}
\affiliation{Frontiers Science Center for Rare Isotopes, and School of Nuclear Science and Technology, Lanzhou University, Lanzhou 730000, China}


\date{\today}

\begin{abstract}
We report the first determination of the Cabibbo-Kobayashi-Maskawa matrix elements \( |V_{cs}| \) and \( |V_{cd}| \) from a global fit to data from inclusive and sum-of-exclusive charm decays. Simultaneously, the heavy quark expansion parameters are determined, and they are in good agreement with results from the literature, validating the robustness of this work. With the current precision, our determined value for \( |V_{cs}| \) is consistent with the world-average value  extracted from exclusive charm decay processes, while a tension of approximately $3\sigma$ is observed for \( |V_{cd}| \) when compared to its exclusive world-average counterpart.
\end{abstract}

\maketitle


\textit{Introduction.} --- Precise tests of the Cabibbo-Kobayashi-Maskawa (CKM) matrix elements, which are fundamental parameters of the Standard Model (SM) of particle physics, are crucial for uncovering dynamics beyond the SM. In this regard, the long-standing tensions in the extracted values of $V_{ub}$ and $V_{cb}$ from exclusive and inclusive $B$ meson decay processes~\cite{HeavyFlavorAveragingGroupHFLAV:2024ctg}—often referred to as the ``$V_{ub,cb}$ puzzles"—have hinted at the presence of non-standard dynamics~\cite{Carvunis:2025vab,Vos:2024qtd,Fael:2024rys,Gambino:2020jvv}. To verify whether these hints are real or merely a consequence of underestimated experimental or theoretical systematic uncertainties, it is imperative to conduct an analogous comparison of the $V_{cs}$ and $V_{cd}$ values obtained from exclusive versus inclusive $D$ meson decays.

The magnitudes of $V_{cs,cd}$ have been determined with high precision through exclusive leptonic and semi-leptonic charmed-hadron decays~\cite{Friday:2025gpj, Pan:2025ajv,BESIII:2024njj, BESIII:2024lnh, BESIII:2023gbn, BESIII:2023wgr, FermilabLattice:2022gku}. However, they have never been measured in inclusive charm decays because experimentally distinguishing strange and non-strange final states in inclusive $D\to X\ell^+\nu$ decays is inaccessible. In this work, we present the first extraction of inclusive values for $|V_{cs,cd}|$, which are obtained by combining the inclusive and sum-of-exclusive $D\to X\ell^+\nu$ data.

Several global fits with different data selection strategies  were performed  to determine the CKM matrix elements. The strategy with  combining the inclusive $D\to X\ell^+\nu$ spectra and the sum of exclusive strange decay rates provides the most optimal fit to the theoretical formulas based on the heavy quark expansion.
Utilizing this method, the inclusive values for $V_{cs,cd}$ are determined to be $|V_{cs}|=0.968 \pm 0.022 \pm 0.026\pm 0.014$ and $|V_{cd}|=0.299 \pm 0.025 \pm 0.007\pm 0.002$. Our result for $|V_{cs}|$ is in perfect consistency with its exclusive counterpart, while a significant discrepancy of approximately $3\sigma$ is observed for $|V_{cd}|$. This finding suggests that the CKM puzzles extend beyond the bottom sector and into the charm sector. Should this discrepancy be confirmed with future increases in precision, it would necessitate a careful re-evaluation of the theoretical frameworks and experimental methodologies employed, or would require new physics models to simultaneously account for the discrepancies observed in both the bottom and charm sectors.

In the remainder of this letter, we first briefly introduce the theoretical formulas for inclusive charm decays. Next, we describe the three strategies used for data selection, the fitting procedure to the theoretical formulas, and the relevant results. Finally, we discuss the physical implications of our findings and conclude.

\textit{Theoretical Formulas.} --- The inclusive decay widths and other inclusive observables can be derived within the framework of operator product expansion in the heavy quark effective theory (HQET). A comprehensive theoretical study for $D$ meson decays is performed in~\cite{King:2021xqp}, presenting analytical results for the electronic inclusive $D$ meson decay widths and the electron moments $\langle E_{e}^{n} \rangle$ up to contributions from dimension-six HQET operators in the leading order of the strong coupling $\alpha_s$. A recent study~\cite{Shao:2025vhe} incorporate the next-to-leading and next-to-next-to-leading order corrections~\cite{Chen:2023osm} to the leading power contributions to these observables, and demonstrates that the both the heavy quark expansion and perturbative strong coupling expansion are well-behaved for semileptonic inclusive $D$ meson decays.

Following~\cite{Shao:2025vhe} and using the same notation, we quote the inclusive decay widths of the $D_i(i=u,d,s)$ mesons,
\begin{widetext}
\begin{align}\label{eq:decaywidth}
\Gamma_{D_{i}\to X_q}=\hat{\Gamma}_0\left|V_{cq}\right|^2 m_c^5 \Big\{ 1 &+\frac{\alpha_s(\mu)}{\pi} {2\over3}\left(\frac{25}{4}- {\pi^2}\right) + \frac{\alpha_s^2(\mu)}{\pi^2}\left[  {\beta_0\over 4} \frac{2}{3} \left(\frac{25}{4}-\pi ^2\right)\log \left(\frac{\mu^2}{m_c^2}\right) +2.14690 n_l-29.88311 \right]  \nonumber \\ 
  & -8\rho\delta_{sq} -\frac{1}{2}\frac{\mu_{\pi}^{2}(D_{i})}{m_{c}^2}-\frac{3}{2}\frac{\mu_{G}^{2}(D_{i})}{m_{c}^{2}}+\left(6+8 \log \left(\frac{\mu^2}{m_c^2}\right)\right) \frac{\rho_D^3\left(D_i\right)}{m_c^3}+\frac{\tau_0(D_i\to X_q)}{m_c^3}+ ...\Big\},
\end{align}
\end{widetext}
where the prefactor is defined as $\hat{\Gamma}_0 = G_F^2 / (192\pi^3)$, and $V_{cq}$ denotes the relevant CKM matrix element, the leading coefficient of the QCD $\beta$-function $\beta_0 = 11 - 2n_f / 3$, and the mass ratio $\rho \equiv m_s^2 / m_c^2$. We choose to use the active quark number $n_f = 4$ corresponding to the light quark number $n_l = 3$. The parameters $\mu_\pi^2$, $\mu_G^2$, $\rho_D^3$ and $\tau_0$ denote the $D$ meson matrix elements of the kinetic energy operator, the chromomagnetic operator, the Darwin operator and the dimension-six four-quark operators, respectively. Unlike in~\cite{Shao:2025vhe}, where the two $c\to q\ (q=d,s)$ transition contributions are summed, here we list them separately. The formulas for the electron energy moments used in this work are also adopted from (3) of~\cite{Shao:2025vhe}. 

Notify that~\eqref{eq:decaywidth} and the electron energy moment formulas in~\cite{Shao:2025vhe} are expressed in terms of the charm quark pole mass, $m_c$. As analyzed in~\cite{Shao:2025vhe}, the pole scheme results suffer from the renormalon problem and hence exhibit poor convergence behavior, and they need to be transformed to a more theoretically robust scheme. In practice, we adopt the 1S mass scheme, and the relation between the pole and 1S masses can be found in~\cite{Hoang:1998hm,Hoang:1998ng,Hoang:1999zc}. 

\textit{Global Fit.} --- 
The theoretical formulas including~\eqref{eq:decaywidth} and (3) of~\cite{Shao:2025vhe} are parameterized by the hadronic matrix elements of the dimension-five and dimension-six HQET operators, as well as the two CKM matrix elements $|V_{cs}|$ and $|V_{cd}|$. We aim to determine these parameters by fitting the theoretical predictions to experimental data. The available measurements, including the branching ratios of $D\to X_{s+d}e^+\nu$,
\begin{eqnarray}\label{eq:Br}
    B(D^{+}&\to X_{d+s} e^+\nu_{e})=0.1602(32),  \\
    B(D^{0}&\to X_{d+s} e^+\nu_{e})=0.0636(15), \nonumber\\
    B(D_{s}&\to X_{d+s} e^+\nu_{e})=0.0631(14), \nonumber
\end{eqnarray}
and the electron energy spectra~\cite{BESIII:2021duu,CLEO:2009uah}, have been translated into corresponding decay widths and electron energy moments in~\cite{Shao:2025vhe}. However, because all these observables are approximately proportional to $|V_{cs}|^2+|V_{cd}|^2$, it is challenging to individually extract the two CKM matrix elements. To overcome this difficulty, we incorporate all available measurements of exclusive semileptonic decay modes to constrain the branching ratios of the inclusive $D_{u,d,s} \to X_{s} e^+ \nu$ and $D_{u,d,s} \to X_{d} e^+ \nu$ processes. 

Our analysis utilizes the world-average values for each exclusive decay mode provided by the Particle Data Group (PDG)~\cite{ParticleDataGroup:2024cfk} and some recent precise measurements, which are summarized in Table~\ref{tab:exclusivemodes} of Appendix~A. For unmeasured channels, we employ isospin symmetry to relate their branching ratios to those of measured modes. This procedure ensures that basically all dominant decay channels are accounted for. Further details regarding the reconstruction of the inclusive semileptonic branching ratios from these exclusive decay modes are described in Appendix~A.

Assuming the uncertainties of the different exclusive decay channels are uncorrelated, we have obtained the inclusive branching ratios for $D_{u,d,s}\to X_se^+\nu$ and $D_{u,d,s}\to X_de^+\nu$ by summing the exclusive decay modes, with the resulting values and their uncertainties listed in Table~\ref{tab:data}. A comparison of these results with direct inclusive measurements~\eqref{eq:Br} reveals consistency within $1-2\sigma$. This agreement indicates that the exclusive measurements successfully collect basically all considerable decay modes and account for the entirety of the inclusive decay, thereby validating the use of the sum-of-exclusive branching ratios as a reliable proxy for the inclusive decay branching ratios.

\begin{table}[htbp]
\centering
\resizebox{\linewidth}{!}{
\begin{tabular}{cccc}
\toprule
 & $D^{+}$ & $D^{0}$ & $D_s$\\
\hline
$X_{s} e^{+}\nu_e$ &$14.60 \pm 0.16\%$ &$ 5.81 \pm 0.06\%$  & $5.58 \pm 0.14\%$ \\
$X_{d} e^{+}\nu_e$ &$0.96 \pm 0.03\%$ &$ 0.45 \pm 0.01\%$ & $ 0.49 \pm 0.03\%$ \\
\bottomrule
\end{tabular}}
\caption{The branching ratios of semi-leptonic inclusive $D$ decays, derived from the sum of the experimentally well-measured exclusive decay modes.}
\label{tab:data}
\end{table}

Our global fit incorporates the following observables to determine the theoretical parameters, including the inclusive $X_{s+d}$ branching ratios~\eqref{eq:Br}, the electron energy moments listed in (12) of~\cite{Shao:2025vhe}, and the sum-of-exclusive $X_{s}$ and $X_d$ branching ratios provided in Table~\ref{tab:data}. As the number of available observables is insufficient to simultaneously determine all HQE parameters along with the CKM matrix elements, we adopt the vacuum insertion approximation~\cite{Bauer:1986bm} for the four-quark operator contributions, which sets all $\tau(D_i\to X_q)=0$~\cite{King:2021xqp}. This is referred to as Scenario 1 of our fitting procedure. To assess the theoretical uncertainties introduced by this assumption, the fits are also performed under a second scenario. In Scenario 2, the weak annihilation parameters are adopted as $\tau(D_d\to X_{d})=\tau(D_s\to X_{s})=\tau_{\rm val}=-0.11~\GeV^3,\tau(D_u\to X_{d,s})=\tau(D_d\to X_{s})=\tau(D_s\to X_{d})=\tau_{\rm nonval}=0.002~\GeV^3$, which are the central values extracted in a global fit in~\cite{Shao:2025vhe}, based on the assumption that all the valence and non-valence contributions are equal. Furthermore, to evaluate the quality and consistency of the sum-of-exclusive data, we conducted the fits using three different data selection strategies: (\textbf{S1}) all the six values listed in Table~\ref{tab:data} are adapted; (\textbf{S2}) only the three $X_s$ values are included, while the $X_d$ values are excluded; (\textbf{S3}) only the three $X_d$ values are included, while the $X_s$ values are excluded. 

The global fit results of the six setups are collected in Table~\ref{tab:fit}, presenting the $\chi^2$'s per degree of freedom and the extracted values of the CKM matrix elements. Only the \textbf{S2} strategy yields a satisfactory goodness-of-fit in both Scenario 1 and Scenario 2, with $\chi^2$'s per degree of freedom below the threshold for significance. They also pass relevant robustness tests. In contrast, the $\chi^2$'s per degree of freedom for the \textbf{S1} and \textbf{S3} strategies in Scenario 1 are greater than at least 3. Consequently, the results from the \textbf{S2} strategy are chosen as the primary findings of this paper. The determined CKM matrix elements are
\begin{eqnarray}\label{eq:results}
|V_{cs}| &=& 0.968 \pm 0.022 \pm 0.026\pm0.014, \\ 
|V_{cd}| &=& 0.299 \pm 0.025 \pm 0.007\pm 0.002\ (\textbf{S2}), \nonumber
\end{eqnarray}
where the first uncertainty arises from experimental data, the second comes from varying the renormalization scale \( \mu \) from 1 to 2.54 \GeV, and the third is due to the weak annihilation contributions. The extracted HQET parameters read $\mu_{\pi}^2(D^{0,+})=0.08(1)~\rm{GeV}^2,~\mu_{G}^2(D^{0,+})=0.34(4)~\rm{GeV}^2,~\rho_{D}^3(D^{0,+})=-0.003(1)~\rm{GeV}^3,~\rho_{LS}^3(D^{0,+})=0.004(1)~\rm{GeV}^3,~\mu_{\pi}^2(D_{s})=0.10(1)~\rm{GeV}^2,~\mu_{G}^2(D_{s})=0.46(3)~\rm{GeV}^2, ~\rho_{D}^3(D_{s})=-0.004(1)~\rm{GeV}^3, ~\rho_{LS}^3(D_{s})=0.005(1)~\rm{GeV}^3$,  which are perfectly consistent with those reported in~\cite{Shao:2025vhe}, and provide further validation for the current fitting procedure.

\begin{table}[H]
\centering
\resizebox{\linewidth}{!}{
\begin{tabular}{cccc}
\toprule
\textbf{S1} ($X_{s,d}$) & $\chi^2/\text{d.o.f.}$ & $|V_{cs}|$ & $|V_{cd}|$ \\
\midrule
Scenario 1 & 3.33 &$0.970 \pm 0.022 \pm 0.026$ & $0.252 \pm 0.006 \pm 0.006$ \\
Scenario 2 & 1.22 &$0.962 \pm 0.021 \pm 0.025$ & $0.253 \pm 0.006 \pm 0.006$ \\
\toprule
\textbf{S2} ($X_s$) & $\chi^2/\text{d.o.f.}$ & $|V_{cs}|$ & $|V_{cd}|$ \\
\hline
Scenario 1 & 0.33 &$0.968 \pm 0.022 \pm 0.026$ & $0.299 \pm 0.025 \pm 0.007$ \\
Scenario 2 & 0.18 &$0.954 \pm 0.021 \pm 0.025$ & $0.297 \pm 0.025 \pm 0.007$ \\
\toprule
\textbf{S3} ($X_d$) & $\chi^2/\text{d.o.f.}$ & $|V_{cs}|$ & $|V_{cd}|$ \\
\midrule
Scenario 1 & 4.08 &$0.982 \pm 0.023 \pm 0.027$ & $0.253 \pm 0.006 \pm 0.005$ \\
Scenario 2 & 1.24 &$0.974 \pm 0.022 \pm 0.026$ & $0.254 \pm 0.006 \pm 0.006$ \\
\bottomrule
\end{tabular}
}
\caption{The values of $\chi^2$ per degree of freedom and the determined values for $|V_{cs}|$ and $|V_{cd}|$ in the global fits with different scenarios and sum-of-exclusive strategies. The first uncertainty arises from experimental data, while the second comes from varying the renormalization scale \( \mu \) from 1 to 2.54 \GeV.}
\label{tab:fit}
\end{table}

\begin{figure}[t]
    \centering
    \includegraphics[width=\columnwidth]{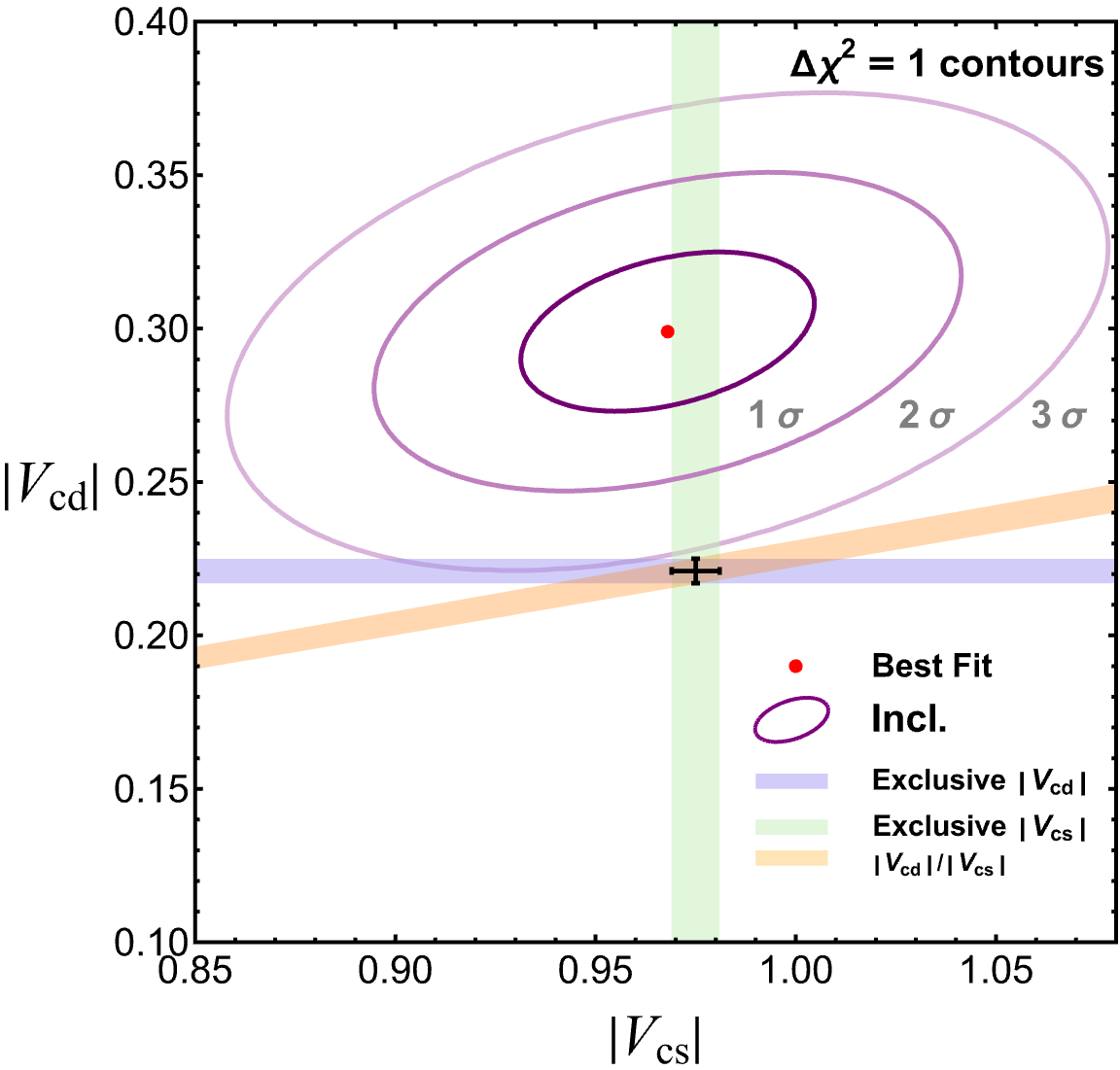}
    \caption{Comparison of inclusive and exclusive determinations of the CKM matrix elements $\left|V_{c s}\right|$ and $\left|V_{c d}\right|$. The dark, light and lighter purple contours represent our global fit results for the inclusive values at the $1 \sigma$, $2 \sigma$ and $3 \sigma$ confidence levels, respectively. The blue and green bands denote the PDG averages of the inclusive determinations at $1 \sigma$, and the orange band indicates the ratio $\left|V_{c d}\right| /\left|V_{c s}\right|$ at $1 \sigma$ confidence level.}
    \label{fig:1}
\end{figure}

A comparison of our determined CKM matrix elements \eqref{eq:results} with the world-average values for $|V_{cs}|$ and $|V_{cd}|$ from PDG~\cite{ParticleDataGroup:2024cfk} is presented in FIG.~\ref{fig:1}. These world-average values shown by the cross in the figure are derived entirely from exclusive decay channels. Our best-fit value for $|V_{cs}|$ is in excellent agreement with its PDG world average, whereas our extracted value for $|V_{cd}|$ exhibits a discrepancy at the level of approximately $3\sigma$ with respect to the PDG average.

As shown in TABLE~\ref{tab:fit}, the \textbf{S1} and \textbf{S3} fitting strategies making use of the $D\to X_d$ data have worse performance than \textbf{S1} strategy. In Scenario 1 of both the \textbf{S1} and \textbf{S3} fitting strategies, the fit quality is very poor. This suggests that the assumption of the sum-of-exclusive $D\to X_d$ branching ratios being complete is likely problematic, pointing to the existence of unobserved $c\to d$ transition channels with significant contributions. This observation is theoretically sound, as the impact of missing decay modes with same amount of branching ratios is more pronounced for the $D\to X_d$ channel, given that its total branching ratio is smaller compared to that of $D\to X_s$. Taking the \textbf{S3} strategy as an example, we further analyze the dependence of the global fit on the $c \to d$ transition data. For the (\textbf{S3}, Scenario 1) setup, as shown in Table~\ref{tab:fit}, the individual contributions to the $\chi^2/\text{d.o.f.}$ from $D^0$, $D^{+}$, $D_s$  are 1.40, 2.12, 0.56, respectively. This specific pattern suggests, from a data-driven perspective, that the existence of undetected exclusive non-strange decay channels is most probable for the $D^{+}$ meson, such as $D^+ \to \pi^+ \pi^- \pi^0 e^+\nu$ from non-$\eta/\omega$ sources and $D^+ \to \pi^+ \pi^- \pi^+ \pi^- e^+ \nu$.

\textit{Summary and Outlook.} --- We  present the first determination of the CKM matrix elements $|V_{cs}|$ and $|V_{cd}|$ using high-precision data from inclusive and sum-of-exclusive charm decays. This work provides a crucial independent test of the CKM mechanism in the charm sector, serving as a benchmark for comparison with results obtained from exclusive charm decay channels. Our final determined values for these CKM matrix elements are $|V_{cs}| = 0.968 \pm 0.022 \pm 0.026\pm 0.014$ and $|V_{cd}| = 0.299 \pm 0.025 \pm 0.007\pm 0.002$. These results reveal a tension of approximately $3\sigma $ with the current world-average exclusive values. This finding, combined with the long-standing puzzles related to $|V_{cb}|$ and $|V_{ub}|$, suggests a possible hint of non-CKM dynamics.


We anticipate that future experiments can provide direct measurements of the sum-of-exclusive $D\to X_s$ and $D\to X_d$ decay branching ratios, which would allow for a better control of their uncertainties. Direct measurements of the inclusive $D\to X_s$ and $D\to X_d$ decay rates would be preferable. This is because they are not subject to the potential issues of missing decay modes and would provide the electron energy spectra, leading to a more robust global fit for determining $|V_{cs}|$ and $|V_{cd}|$. 

\textit{Acknowledgement.} --- We wish to thank Matteo Fael and Wen-Jie Song for enlightening discussions on calculations of electronic energy spectrum, to Ying-Ao Tang for suggestions on the extraction of electronic energy moments, and especially to Dong Xiao for his inspiring discussions on various experimental aspects, respectively. Special thanks go to Long Chen and Yan-Qing Ma for providing the numerical NNLO corrections to the partonic decay widths and the corresponding electron energy moments. This work is supported by Natural Science Foundation of China under grant No.~12375086, 12522506 and 12335003, and by the Fundamental Research Funds for the Central Universities under No. lzujbky-2023-stlt01, lzujbky-2024-oy02 and lzujbky-2025-eyt01.
\appendix

\section{Appendix A: Reconstruction of Inclusive Branching Ratios from Exclusive Data}

\begin{widetext}

\begin{table}[htbp]
\begin{tabular}{lc lc lc}
\toprule
\multicolumn{2}{c}{$D^+$ decays} & 
\multicolumn{2}{c}{$D^0$ decays} & 
\multicolumn{2}{c}{$D_s$ decays} \\
\cmidrule(lr){1-2} \cmidrule(lr){3-4} \cmidrule(lr){5-6}
Mode & BR(\%) & Mode & BR(\%) & Mode & BR(\%) \\
\midrule
$D^+ \to \bar{K}^0 e^+ \nu_e$ & $8.72 \pm 0.09$ & 
$D^0 \to K^- e^+ \nu_e$ & $3.549 \pm 0.026$ & 
$D_s \to \phi e^+ \nu_e$ & $2.34 \pm 0.12$ \\
$(K^-\pi^+)_{\bar{K}^*(892)^0} e^+ \nu_e$ & $3.54 \pm 0.09$~\cite{BESIII:2015hty} & 
$(\bar{K}^0 \pi^-)_{\text{S-wave}} e^+ \nu_e$ & $0.079 \pm 0.017$ & 
$D_s \to \eta e^+ \nu_e$ & $2.27 \pm 0.06$ \\
$(K^-\pi^+)_{\text{S-wave}} e^+ \nu_e$ & $0.228 \pm 0.011$ & 
$D^0 \to K^*(892)^- e^+ \nu_e$ & $2.04 \pm 0.047$~\cite{BESIII:2024xjf}  & 
$D_s \to \eta' e^+ \nu_e$ & $0.81 \pm 0.04$ \\
$D^+ \to \bar{K}_1(1270)^0 e^+ \nu_e$ & $0.230 \pm 0.026$ &
$D^0 \to \bar{K}_1(1270)^- e^+ \nu_e$ & $0.101 \pm 0.018$ &
$D_s \to f_0(980) e' \nu_e$ & $0.164 \pm 0.013$ \\
\midrule
$D^+ \to \eta e^+ \nu_e$ & $0.111 \pm 0.007$ &
$D^0 \to \pi^- e^+ \nu_e$ & $0.291 \pm 0.004$ &
$D_s \to K^0 e^+ \nu_e$ & $0.288 \pm 0.026$ \\
$D^+ \to \pi^0 e^+ \nu_e$ & $0.372 \pm 0.017$ &
$D^0 \to \rho(770)^- e^+ \nu_e$ & $0.145 \pm 0.007$ &
$D_s \to K^{*}(892)^0 e^+ \nu_e$ & $0.205 \pm 0.020$ \\
$D^+ \to \pi^+\pi^- e^+ \nu_e$ & $0.245 \pm 0.008$ &
$D^0 \to a(980)^- e^+ \nu_e$ & $0.0133^{+0.0034}_{-0.0030}$ & & \\
$D^+ \to \pi^0\pi^0 e^+ \nu_e$ & $0.0315 \pm 0.0027$~\cite{BESIII:2018qmf} & & & & \\
$D^+ \to \omega e^+ \nu_e$ & $0.169 \pm 0.011$ & & & & \\
$D^+ \to \eta' e^+ \nu_e$ & $0.020 \pm 0.004$ & & & & \\
$D^+ \to a(980)^0 e^+ \nu_e$ & $0.017 \pm 0.008$ & & & & \\
\bottomrule
\end{tabular}
\caption{Branching ratios of selected exclusive decays of the $D^+$, $D^0$, and $D_s$ mesons in units of percentage. All the values without references are taken from the PDG~\cite{ParticleDataGroup:2024cfk}.}
\label{tab:exclusivemodes}
\end{table}

\end{widetext}

The branching ratios for the inclusive \( D^0, D^+, D_s \to X_{s,d} \ell \nu \) decays are reconstructed by summing over the relevant exclusive decay modes, which are listed in TABLE~\ref{tab:exclusivemodes} with their measured branching ratios. All of these values are taken from the PDG~\cite{ParticleDataGroup:2024cfk}, 
except for three cases. The branching ratio of 
\(D^+ \to (K^-\pi^+)_{\bar{K}^*(892)^0} e^+ \nu_e\) 
is taken from~\cite{BESIII:2015hty}; that of 
\(D^0 \to K^*(892)^- e^+ \nu_e\) is taken from~\cite{BESIII:2024xjf}; 
the branching ratio of \(D^+ \to \pi^0 \pi^0 e^+ \nu\) 
is reconstructed in this work based on the results of~\cite{BESIII:2018qmf}.
Some of them can be summed directly, while the others need to be prefabricated, with the details described below. We note that possible contributions from additional rare exclusive channels are not included in the reconstruction, but we regard that their branching ratios are negligible compared to the quoted modes and can be safely covered by the overall uncertainties at the current stage. 

For the decay \( D^+ \to X_s e^+ \nu \), we consider the measured final $X_s$ states
\[
X_s : \bar{K}^0,\ (K^-\pi^+)_{\bar{K}^*(892)^0},\ (K^-\pi^+)_{\text{S-wave}},\ \bar{K}_1(1270)^0\;.
\]
In order to obtain \(\mathcal{B}(D^+ \to \bar{K}^*(892)^0 e^+ \nu)\),
the branching ratio \(\mathcal{B}(D^+ \to K^-\pi^+ e^+ \nu)_{\bar{K}^*(892)^0}\) is rescaled by dividing it by the isospin-derived branching ratio \(\mathcal{B}(\bar{K}^{*0} \to K^- \pi^+) = 2/3\). Similarly, the $S$-wave component is rescaled by the same factor to account for the full isospin multiplet.

For the decay \( D^+ \to X_d e^+ \nu \), we consider 
\[
X_d : \eta,\ \pi^0,\ (\pi^+\pi^-),\ (\pi^0\pi^0),\ \omega,\ \eta' ,\ a_0(980)^0 .
\]
In the reconstruction, we remove the \(\omega \to \pi^+\pi^-\) contribution from the $D^+\to \omega e^+\nu$ decay, to avoid double counting with the \(\pi^+\pi^-\) channel. As the $(\pi^0\pi^0)$ branching ratio is not available in PDG, we estimate it in the following way. The statistical property of identical particles and discrete symmetries forbid $(\pi^0\pi^0)$ decaying from pseudoscalar and vector resonances, and we regard \( f_0(500) \to \pi^0 \pi^0 \) to be the dominant source of $(\pi^0\pi^0)$. According to~\cite{BESIII:2018qmf}, the branching ratio for \( f_0(500) \to \pi^0 \pi^0 \) is half of that for \( f_0(500) \to \pi^+ \pi^- \) due to the constraints from isospin symmetry and the additional factor of \( 1/2 \) arising from the fact that \( \pi^0 \pi^0 \) involves identical particles. Thus, the branching ratio for the \(\pi^0\pi^0\) channel is reconstructed to be half of the measured result \(\mathcal{B}(D^+ \to f_0(500) e^+ \nu_e,f_0(500)\to \pi^+\pi^-)=(6.30\pm 0.43\pm 0.32) \times 10^{-4}\)~\cite{BESIII:2018qmf}.

For the decay \(D^0 \to X_s e^+ \nu\), we take into account 
\[
X_s: K^- \ ,(\bar{K}^0\pi^-)_{\text{S-wave}}\ ,K^*(892)^-\ ,\bar{K}_1(1270)^-.
\]
Assuming the isospin symmetry, the branching ratio for S-wave contribution is rescaled by dividing by \(2/3\) for the full isospin multiplet.

For the decay \(D^0 \to X_d e^+ \nu\), we take into account 
\[
X_d: \pi^- \ , \rho(770)^- \ , a(980)^-.
\]

For the decay \(D_s \to X_s e^+ \nu\), we consider 
\[
X_s:\phi \ ,\eta \ , \eta'\ , f_0(980) .
\]

For the decay \(D_s \to X_d e^+ \nu\), we consider 
\[
X_d: K^0 \ , K^*(892)^0.
\]


\end{document}